\documentclass{sig-alternate-2013}

\usepackage{url}
\usepackage{multirow}
\usepackage{times}

\newfont{\mycrnotice}{ptmr8t at 7pt}
\newfont{\myconfname}{ptmri8t at 7pt}
\let\crnotice\mycrnotice%
\let\confname\myconfname%

\permission{Permission to make digital or hard copies of all or part of this work for personal or classroom use is granted without fee provided that copies are not made or distributed for profit or commercial advantage and that copies bear this notice and the full citation on the first page. Copyrights for components of this work owned by others than ACM must be honored. Abstracting with credit is permitted. To copy otherwise, or republish, to post on servers or to redistribute to lists, requires prior specific permission and/or a fee. Request permissions from permissions@acm.org.}
\conferenceinfo{KDD'13,}{August 11--14, 2013, Chicago, Illinois, USA.}
\CopyrightYear{2013} 
\crdata{978-1-4503-2174-7/13/08} 
\begin{document}

\title{Modeling social network evolution with user activity dynamics}
\title{Coupling user activity and structure evolution: a new model of social networks}
\title{Efficient information diffusion \\and the evolution of social networks}
\title{Diverse Link Creation Mechanisms \\in the Evolution of Social Networks}
\title{User Activity and Link Creation \\in the Evolution of Social Networks}
\title{Does Traffic Matter?\\User Activity and Link Creation in Online Social Networks}
\title{Why does Alice follow Bob?\\User Activity and Traffic in Online Social Networks}
\title{User Activity, Traffic, and Link Creation in Information Diffusion Networks}
\title{User Activity, Traffic, and Link Creation in Online Social Networks}
\title{The Role of Information Diffusion \\ in the Evolution of Social Networks}

\numberofauthors{1}
\author{
\alignauthor Lilian Weng$^1$, Jacob Ratkiewicz$^2$, Nicola Perra$^3$, Bruno Gon\c{c}alves$^4$, Carlos Castillo$^5$, Francesco Bonchi$^6$, Rossano Schifanella$^7$, Filippo Menczer$^1$, Alessandro Flammini$^1$\\
\affaddr{$^1$School of Informatics and Computing, Indiana University Bloomington, USA $^2$Google Inc.}\\
\affaddr{$^3$Laboratory for the Modeling of Biological and Socio-technical Systems, Northeastern University, USA}\\
\affaddr{$^4$Aix Marseille Universit\'{e}, CNRS, CPT, UMR 7332, Marseille, France}\\
\affaddr{$^5$Qatar Computing Research Institute $^6$Yahoo! Research Barcelona}\\
\affaddr{$^7$Department of Computer Science, University of Torino, Italy}\\
}

\maketitle

\begin{abstract}

Every day millions of users are connected through online social networks, generating a rich trove of data that allows us to study the mechanisms behind human interactions. Triadic closure has been treated as the major mechanism for creating social links: if Alice follows Bob and Bob follows Charlie, Alice will follow Charlie.
Here we present an analysis of longitudinal micro-blogging data, revealing a more nuanced view of the strategies employed by users when expanding their social circles. While the network structure affects the spread of information among users, the network is in turn shaped by this communication activity. This suggests a link creation mechanism whereby Alice is more likely to follow Charlie after seeing many messages by Charlie. 
We characterize users with a set of parameters associated with different link creation strategies, estimated by a Maximum-Likelihood approach. 
Triadic closure does have a strong effect on link formation, but shortcuts based on traffic are another key factor in interpreting network evolution. However, individual strategies for following other users are highly heterogeneous. Link creation behaviors can be summarized by classifying users in different categories with distinct structural and behavioral characteristics. Users who are popular, active, and influential tend to create traffic-based shortcuts, making the information diffusion process more efficient in the network. 

\end{abstract}




\section{Introduction}

\begin{figure}
\centering
\includegraphics[width=0.9\columnwidth]{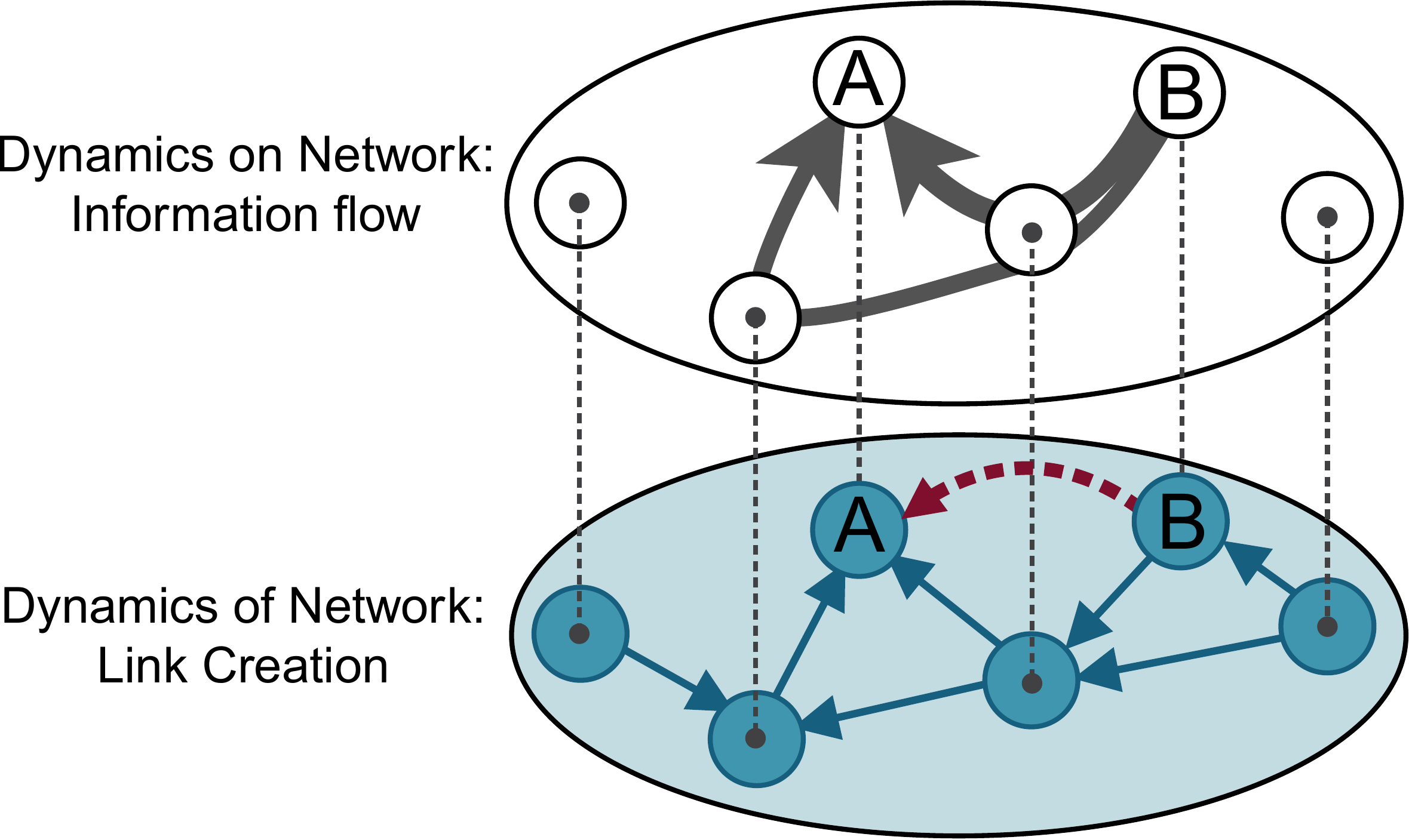}
\caption{The dynamics \emph{of} and \emph{on} the network are strongly coupled. The bottom layer illustrates the social network structure,  where the blue arrows represent ``follow'' relationships with the direction of information flow. The dashed red arrow marks a newly created link. 
The upper layer depicts the flow of information between people in the same group, leading to the creation of the new link.
}
\label{fig:illustration}
\end{figure}

User activity in online social networks is exploding. Social and micro-blogging networks such as Facebook, Twitter, and Google Plus every day host the information sharing activity of billions of users. Using these systems, people communicate ideas, opinions, videos, and photos among their circles of friends and followers across the world. These interactions generate an unprecedented amount of data that can be used as a social observatory, providing a unique opportunity to shed light on the mechanisms of human communication with a quantitative approach~\cite{lazer-09,cho-09,kumar-06,angelina, vespignani2009, vespignani2011modelling}.

Research on social media revolves around two main themes: communication and its social network substrate. Most network models focus on either the structural growth of the system --- the dynamics \emph{of} the network --- or information diffusion processes --- the dynamics \emph{on} the network. The present work establishes a feedback loop between these two dynamics. 

Much effort has been devoted to modeling the evolution of social networks~\cite{Wasserman-F, Alexlibro, Newmanlibro, angelina}. Among proposed mechanisms of how a link is created, \emph{triadic closure}~\cite{simmel, Granovetter} is a simple but powerful principle to model the evolution of social networks based on shared friends: two individuals with mutual friends have a higher than random chance to establish a link.
In directed networks, such as Twitter or Yahoo! Meme, triadic closure implies a particular order with respect to the direction of links: once Alice follows Bob and Bob follows Charlie, Alice will follow Charlie.
Triadic closure has been observed in both undirected and directed online social networks and incorporated into several network growth models~\cite{Leskovec-08,Krackhardt,romero2010closure}. However, most existing models do not take user activity --- or how information spreads on the network --- into consideration.

Social micro-blogging networks, such as Twitter, Google Plus, Sina Weibo, and Yahoo! Meme, are designed for information sharing. As illustrated in Fig.~\ref{fig:illustration}, the social network structure constrains communication patterns, but information propagated through the network also affect how agents behave and ultimately how the network changes and grows. In this paper we study the role of information diffusion in shaping the evolution of the network structure, and the individual strategies that bring about this effect by way of creating social links.

The major contribution of this paper is to present clear evidence that information diffusion affects network evolution at both system-wide and individual levels. 
In particular, we find that a considerable portion of new links are \emph{shortcuts based on information flow} (\S~\ref{sec:4}). 
There is significant statistical evidence for triadic closure as a link creation mechanism, but also that users tend to link to people who have generated content they have seen (\S~\ref{sec:4.1}).
Furthermore, not all users apply the same strategy to grow their social connections; users with high in-degree tend to pay more attention to traffic (\S~\ref{sec:4.2}). 
However, shortcuts are not equally probable; we find that users follow the most active sources of content; purely topological mechanisms cannot account for these shortcuts (\S~\ref{sec:4.3}).
As a result, traffic-based shortcuts can make the social network more efficient in terms of information diffusion (\S~\ref{sec:4.4}).
In \S~\ref{sec:5}, we perform a Maximum Likelihood Estimation analysis to quantify the system-wide prevalence of different link creation strategies.
Finally, the categorization of users suggests the existence of several distinct link formation behaviors (\S~\ref{sec:6}).
Our findings identify information diffusion dynamics as a key factor in the evolution of social networks.

\section{Background}


Early models concerning communication dynamics were inspired by studies of epidemics, assuming that a piece of information could pass from one individual to another through social contacts~\cite{rapoport53,goffman64,daley64,bailey75,SIR}. These models have been extended to include cascade phenomena~\cite{goldenberg01}, factors that influence the speed of spreading such as information recency~\cite{moreno04}, the heterogeneity in connectivity patterns~\cite{Romu01-1}, clustering~\cite{onnela06-1}, user-created content~\cite{bakshy09}, and temporal connectivity patterns~\cite{morris95-1,butts08-1,butts09-1,perra12-1}. 
An alternative class of models is based on the idea of a threshold; you propagate an idea when some number of friends communicate it to you~\cite{threshold_model,Morris00}. These models are believed to be relevant in the diffusion of rumors, norms, and behaviors~\cite{centola2010}, and have been extended to study the role of competition for finite attention~\cite{Weng:2012scirep}. 
The large majority of these studies consider either a \emph{static} or \emph{annealed} underlying social network, under the assumption that the network evolves on a longer (slower) time scale than the information spread. 
Recent research has addressed the modeling of intermediate cases, in which the two time scales are comparable. These approaches consider the two dynamics as either independent~\cite{Rocha:2010,perra12-2} or coupled~\cite{volz09-1,schwartz10-2}. The foundations of this last class of models are very similar to those explored in this paper. However, thus far, these models have focused mainly on epidemic processes in which links are deleted or rewired according to the disease status of each node~\cite{volz09-1,schwartz10-2}. The social systems considered in this paper are governed by quite different underlying mechanisms.

Models devoted to reproducing the growth and evolution of network topology have traditionally focused on defining basic mechanisms driving link creation~\cite{Wasserman-F,Newmanlibro,Alexlibro}. From the first model proposed in 1959 by  Erd\"os and R\'enyi~\cite{Erdos}, many others have been introduced capturing different properties observed in real networks, such as the small-world phenomenon~\cite{Watts-S,Leskovec-08,Krackhardt,romero2010closure}, large clustering coefficient~\cite{Watts-S,Leskovec-08,Krackhardt,romero2010closure}, temporal dynamics~\cite{perra12-1,Rocha:2010}, information propagation~\cite{barbieri2013cascade}, and heterogeneous distributions in connectivity patterns~\cite{BA, Kleinberg-K-R-R, kumar2000stochastic, Krapivsky-R,Dorogo-M-S00a, santo-PRL}. In particular, this latter property was first described by the preferential attachment~\cite{BA} and copy models~\cite{kumar2000stochastic}.

In the social context, the rationale behind preferential attachment mechanisms is that people prefer linking to well connected individuals~\cite{BA}. Although very popular, this prescription alone is not sufficient to reproduce other important features of social networks. Other models have been put forth to fill this gap, including ingredients such as homophily~\cite{holme2006opinion, McPherson2001homophily, papa2012similarity, Aiello2011Friendship,gallos2012people} and triadic closure~\cite{simmel, Granovetter,Leskovec-08,Krackhardt,romero2010closure}. 
%
Homophily describes the tendency of people to connect with others sharing similar features~\cite{McPherson2001homophily,holme2006opinion}. Its impact on link creation in large-scale online networks is a recent topic of discussion~\cite{papa2012similarity, Aiello2011Friendship,gallos2012people}. 
%
The triadic closure mechanism is based on the intuition that two individuals with mutual friends have a higher probability to establish a link~\cite{simmel, Granovetter}. This tendency has been observed in both undirected and directed online social networks and incorporated into several network growth models~\cite{Leskovec-08,Krackhardt,romero2010closure}. In particular Leskovec \textit{et al.} have tested triadic closure against many other mechanisms in four different large-scale social networks~\cite{Leskovec-08}. By using Maximum Likelihood Estimation (MLE)~\cite{cowan} they have identified triadic closure as the best rule, among those considered, to explain link creation.

Link prediction algorithms, aimed at inferring new connection that may happen in the near future given a current snapshot of the network structure, could be used as ingredients for modeling network evolution.
Common approaches consider link prediction as a classification task or ranking problem, using node similarity~\cite{LibenNowell2007LinkPred,katz}, the hierarchical structure of the network~\cite{Clauset:2008fk}, random walks~\cite{backstrom2011supervised}, graphical models~\cite{lou2010},  and user profile features~\cite{Schifanella20120folk}.

Although similar in spirit, our approach is different from this large body of literature. We do not consider link prediction, nor agent based simulations in which the structural behavior of each user is modeled by a set of rules. We adopt the MLE framework extending the work of Leskovec \textit{et al.}~\cite{Leskovec-08}. We extend the notion of triadic closure by considering mechanisms based on traffic, or more in general, users activity. We explicitly study the coupling between the dynamics \emph{of} and \emph{on} the network, connecting these two previously separated themes of research, in the context of online social networks.

\section{Meme Dateset}
\label{sec:3}

We study \emph{Yahoo! Meme,} a social micro-blogging system similar to Twitter, which was active between 2009 and 2012. We have access to the entire history of the system, including full records of every message propagation and link creation event, from April 2009 until March 2010.
%
A user $j$ following a user $i$ is represented in the follower network by a directed edge $\ell = (i,j)$, indicating $j$ can receive messages posted by $i$. We adopt this notation, in which the link creator is the target, to emphasize the direction of information flow.
Edges are directed to account for asymmetric relations between users; a node can follow another without being followed back. In our notation, the in-degree of a node $i$ is the number of people followed by $i$, and the out-degree is the number of $i$'s followers. Users can repost received messages, which become visible to their followers. When user $j$ reposts content from $i$, we infer a flow of information from $i$ to $j$. Each link is weighted by the numbers of messages from $i$ that are reposted or seen by $j$.

\begin{figure}[t]
\centering
\includegraphics[width=0.85\columnwidth]{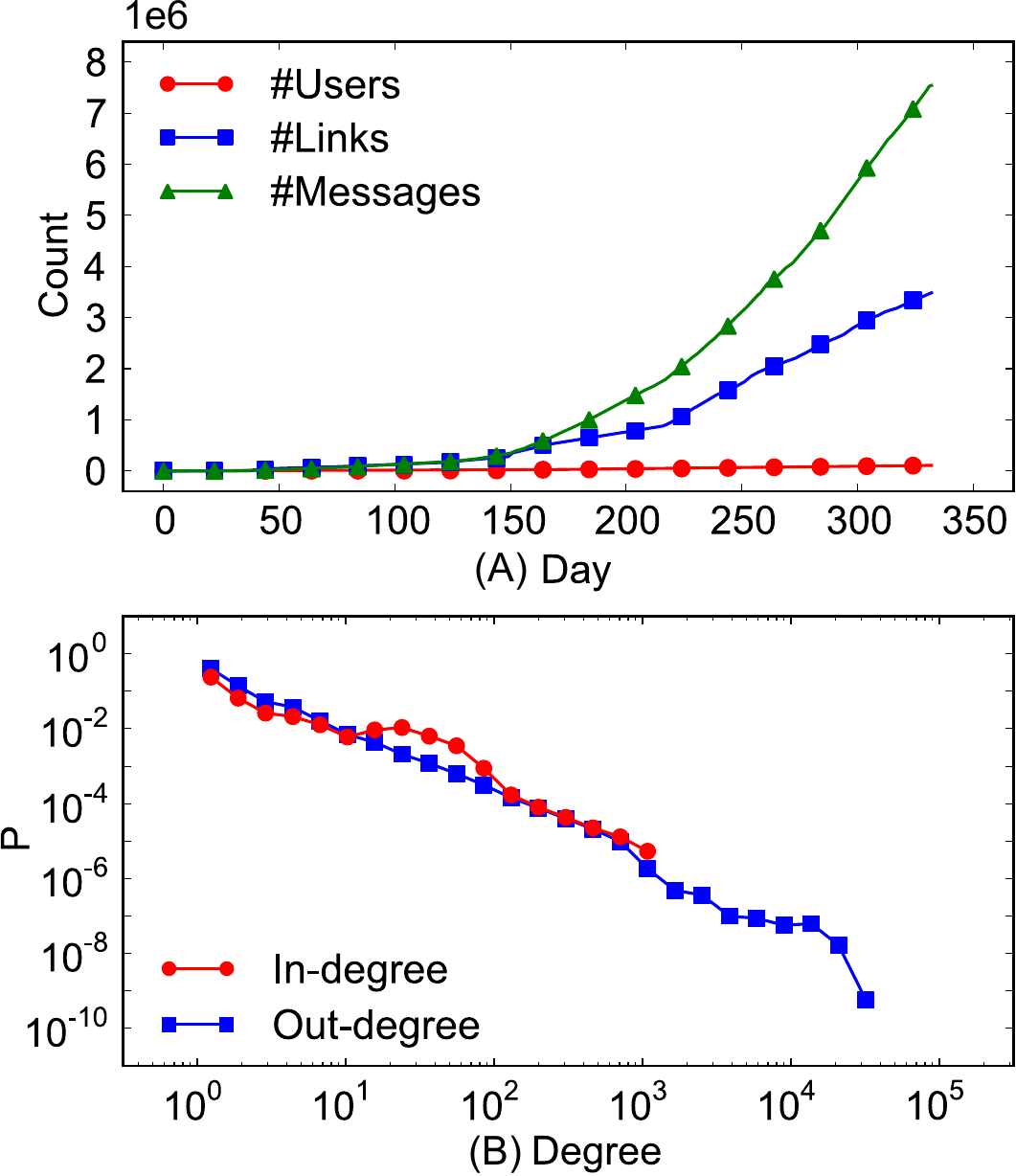}
\vspace{-1em}
\caption{General statistics of the Yahoo! Meme system.
(A)~The growth of the system in time, the number of users (red circles), links (blue squares) and messages (green triangles).
(B)~Broad distributions of in-degree and out-degree in the follower network of Yahoo! Meme. Users were not allowed to follow more than 1,000 people, which is the maximum in-degree a node can attain. 
\vspace{-1em}
}
\label{fig:stats}
\end{figure}

At the end of the observation period, the Yahoo! Meme follower network consisted of 128,199 users with at least one edge, connected by a total of 3,485,361 directed edges. Fig.~\ref{fig:stats} displays general statistics about the growth and structure of the network. 

\section{Link Creation Mechanisms}
\label{sec:4}

\begin{figure}
\centering
\includegraphics[width=0.85\columnwidth]{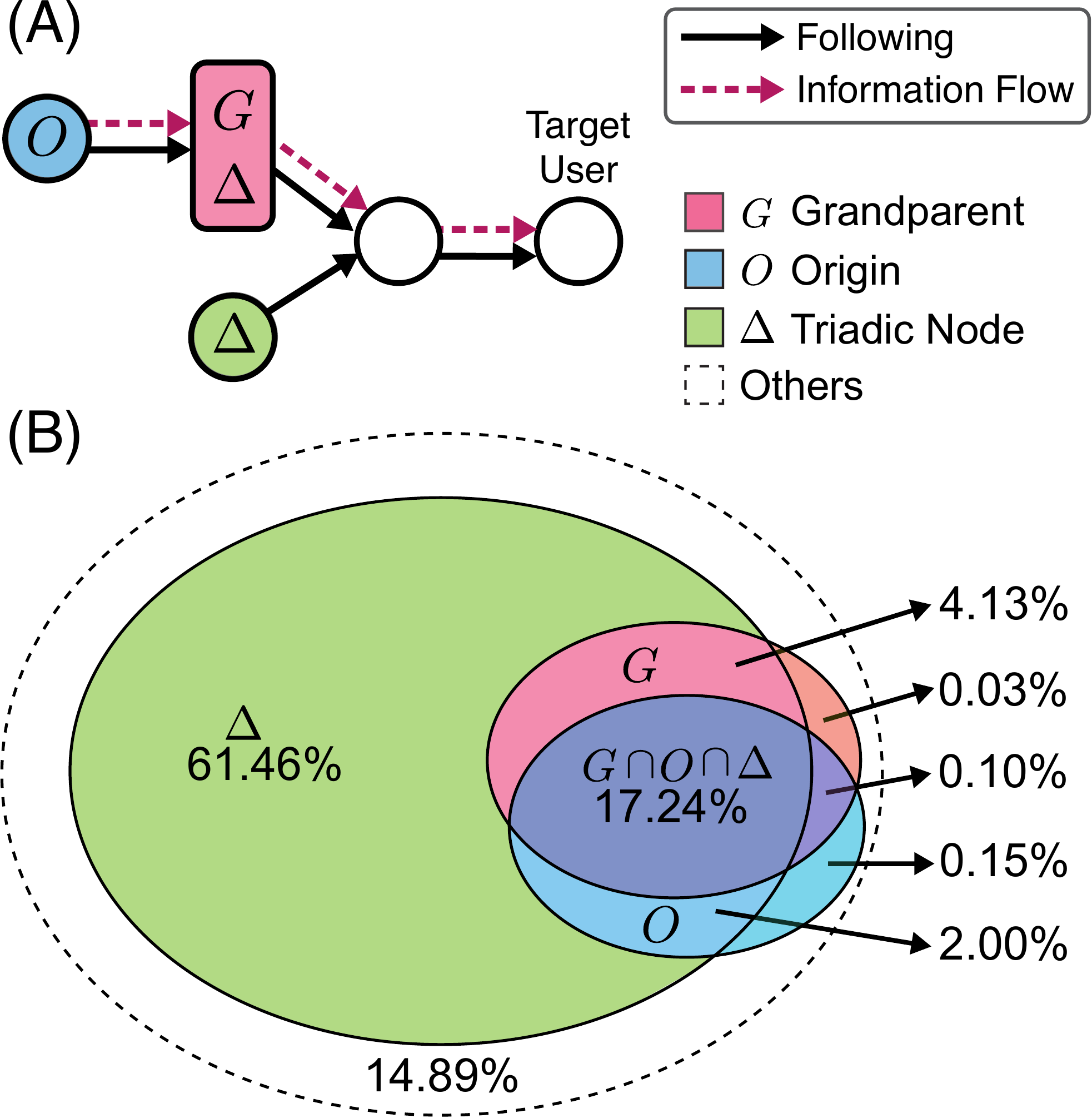}
\vspace{-1em}
\caption{
(A)~Illustration of the link creation mechanisms.
(B)~Venn digram of the proportions of grandparent, origin, and triadic closure links among all existing edges.
\vspace{-1em}}
\label{fig:definition}
\end{figure}


When users post or repost messages, all their followers can see these posts and might decide to repost them, generating paths that together form cascade networks. When receiving a reposted message, a user in such a path can see both the \emph{grandparent} ($G$, the user two steps ahead in the path) and the \emph{origin} ($O$, original source). 
A user may decide to follow a grandparent or origin, receiving their future messages directly. These new links create \emph{shortcuts} connecting users at any distance in the network. 
%
%
A triadic closure occurs when a user follows a \emph{triadic node} ($\Delta$, the user two steps away in the follower network).
The definitions of different types of link creation mechanisms are illustrated in Fig.~\ref{fig:definition}(A).

The Venn diagram in Fig.~\ref{fig:definition}(B) shows the proportions of links of different types and the logical relationships between these sets of links. 
We observe that 84.8\% of new edges consist of triadic closures, 21.5\% form shortcuts to grandparent, and 19.5\% to origins. 
Note that not all the grandparents are triadic nodes, because users are allowed to repost messages from people they are not following in Yahoo! Meme. This account for 0.03\% of links.
There is a large overlap between triadic closure links and traffic-based shortcuts. This can be explained by the phenomenon that most real-world information cascades are shallow~\cite{bakshy2011everyone} and thus triadic closure links and traffic-based shortcuts coincide.

This evidence suggests that traffic-based link creation mechanisms are an important complement to the triadic closure in modeling network evolution. Actions of posting and reposting induce the creation of shortcuts, shaping the structure of the network. Newly created links in turn determine what messages are seen by users, making the network more efficient at spreading information.

\subsection{Statistical Analyses of Shortcuts} 
\label{sec:4.1}

To quantify the statistical tendency of users to create shortcuts, let us consider every single link creation in the data as an independent event. We test the null hypotheses that links to grandparents, origins, and triadic nodes are generated by choosing targets at random among the users not already followed by the creator.

We label each link $\ell$ by its creation order, $1 \leq \ell \leq L$, where $L$ is the total number of links. For each link, we can compute the likelihood of following a grandparent by chance:
\[
p_G(\ell)=\frac{N_G(\ell)}{N(\ell) - k(\ell)-1}
\]
where $N_G(\ell)$ is the number of distinct grandparents seen by the creator of $\ell$ at the moment when $\ell$ is about to be created; $N(\ell)$ is the number of available users in the system when $\ell$ is to be created; $k(\ell)$ is the in-degree of  $\ell$'s creator at the same moment; and the denominator is the number of potential candidates to be followed. 
The indicator function for each link $\ell$ denotes whether the link connects with a grandparent or not in the real data:
\begin{align*} 
\mathbf{1}_G (\ell)&= \left\{ 
  \begin{array}{l l}
    1 & \text{if $\ell$ links to a grandparent} \\
    0 & \text{otherwise.} \\
  \end{array}\right.
\end{align*}
The expected number of links to grandparents according to the null hypothesis can be then computed as:
\[
E_G=\sum_{\ell=1}^{L} p_G(\ell)
\]
and its variance is given by:
\[
\sigma^2_G = \sum_{\ell=1}^{L} p_{G}\left(\ell \right)\left(1-p_{G}\left(\ell \right)\right)
\]
while the corresponding empirical number is: 
\[
S_G= \sum_{\ell=1}^{L} \mathbf{1}_G(\ell).
\]
%
According to the Lyapunov central limit theorem,\footnote{Lyapunov's condition, $\frac{1}{\sigma_{n}^4} \sum_{\ell=1}^n E[(X(\ell) - p(\ell))^4] \overset{n\to\infty}{\longrightarrow} 0$  where $X(\ell)$ is a random Bernoulli variable with success probability $p(\ell)$~\cite{billingsleyprobability}, is consistent with numerical tests. Details are omitted for brevity.} the variable $z_G=(S_G - E_G)/\sigma_G$ is distributed according to a standard normal $\mathcal{N}(0,1)$. For linking to origins ($O$) or triadic nodes ($\Delta$), we can define $z_O$ and $z_{\Delta}$ similarly.
%
In all three cases, using a $z$-test, we can reject the null hypotheses with high confidence ($p < 10^{-10}$).
We conclude that links established by following grandparents, origins or triadic nodes happen much more frequently than by random connection. These link creation mechanisms have important roles in the evolution of the social network.

\subsection{User Preference}
\label{sec:4.2}

\begin{figure}
\centering
\includegraphics[width=0.925\columnwidth]{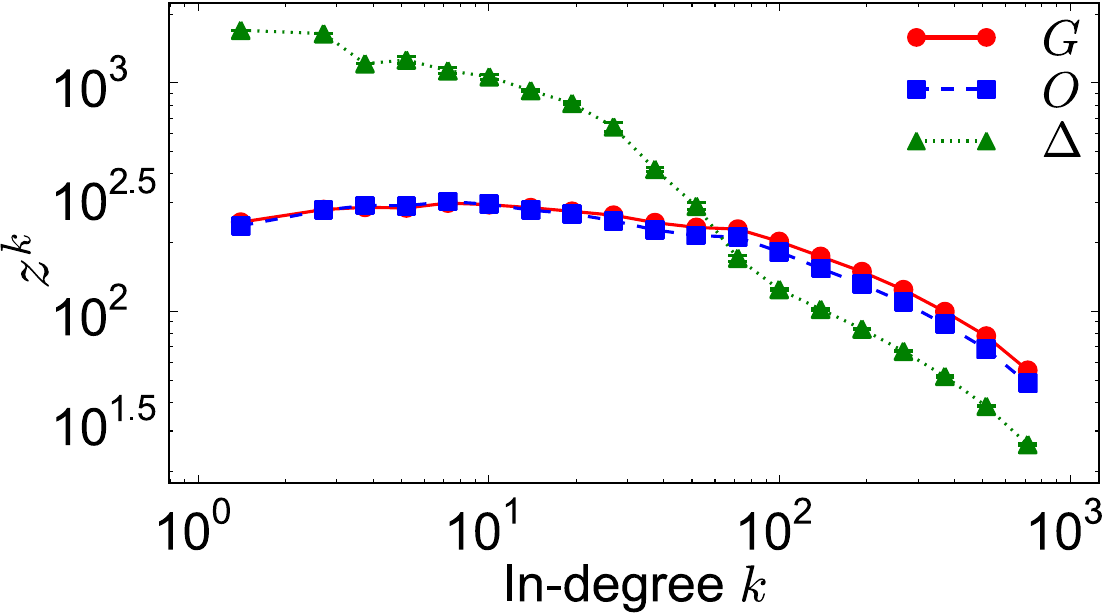}
\caption{Individual preferences for following grandparents (red circles), origins (blue squares) and triadic nodes (green triangles) change with the in-degree of the link creator.}
\label{fig:preference}
\end{figure}

To study the dependence of the link formation tendencies on the different stages of an individual's lifetime, let us compute $z^k_G$, $z^k_O$ and $z^k_{\Delta}$ for links created by users with in-degree $k$, that is, those who are following $k$ users at the time when the link is created.
Fig.~\ref{fig:preference} shows that the principle of triadic closure dominates user behavior when one follows a small number of users ($k < 75$). In the early stages, one does not receive much traffic, so it is natural to follow people based on local social circles, consistently with triadic closure.
However, users who have been active for a long time and have followed many people ($k > 75$) have more channels through which they monitor traffic. This creates an opportunity to follow others from whom they have seen messages in the past. 

\subsection{Traffic Bias}
\label{sec:4.3}


\begin{figure}[t]
\centering
\includegraphics[width=0.925\columnwidth]{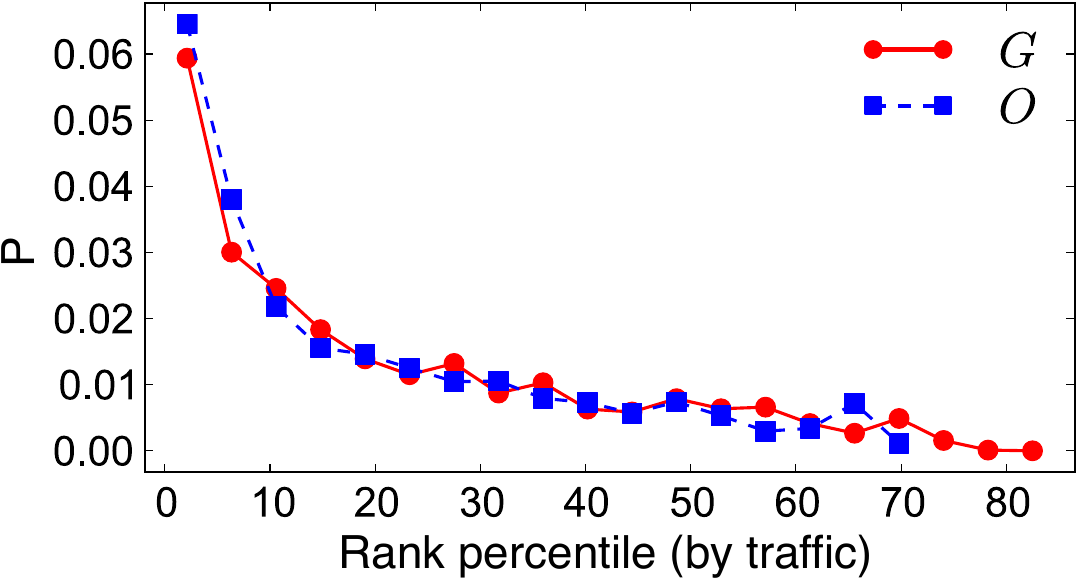}
\caption{
Probability density of followed grandparents (red circles) or origins (blue squares) having a certain rank percentile. 
Link targets are ranked so that the link creator has seen more messages from a user with smaller rank percentile.
}
\label{fig:weights}
\end{figure}

Further inspection of the empirical data reveals that not all shortcuts are equally likely; users tend to follow those who have often been sources of seen messages. To investigate this, consider all new shortcuts to grandparents or origins. For each shortcut, we rank all the available grandparent or origin candidates according to how many of their messages have been seen by the creator prior to the link formation.
We plot the probability of a followed grandparent or origin having a certain rank percentile in Fig.~\ref{fig:weights}. The plot clearly demonstrates that repeated exposure to contents posted by a user increases the probability of following that user. This is analogous to the way in which we are more likely to adopt a piece of information or behavior to which we are exposed multiple times~\cite{bakshy09, centola2010, romero_differences_2011, Hodas2012_attention}. 
%
%
This observation shows that topology alone is insufficient to explain the evolution of the network; activity patterns --- the dynamics \emph{on} the network --- are a necessary ingredient in describing the formation of new links.

\subsection{Link Efficiency}
\label{sec:4.4}

\begin{figure}[!t]
\centering
\includegraphics[width=0.9\columnwidth]{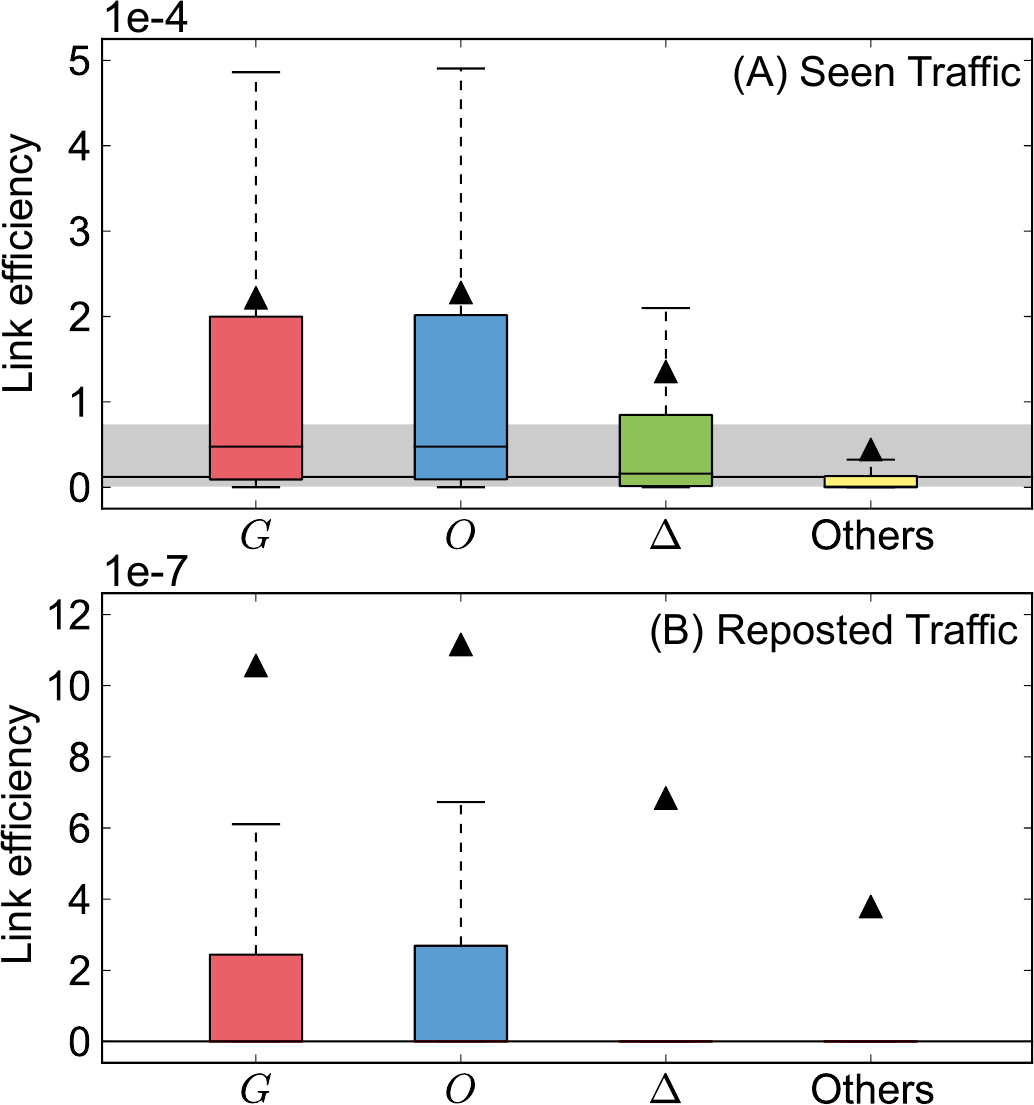}
\vspace{-1em}
\caption{Efficiency of links created according to different mechanisms, or average number of messages (A)~seen or (B)~reposted per time unit. 
Each box shows data within lower and upper quartile. Whiskers represent the 99th percentile. The triangle and line in a box represent the median and mean, respectively. Note that the mean can fall outside the shown quantiles for skewed distributions.
The grey area and the black line across the entire figure mark the interquartile range and the median of the measure across all links, respectively.
}
\label{fig:eff}
\end{figure}

In information diffusion networks like Twitter and Yahoo! Meme, social links may have a key efficiency function of shortening the distance between information creators and consumers. An efficient link should be able to convey more information to the follower than others. Hence we define the \emph{efficiency} of link $\ell$ as the average number of posts seen or reposted through $\ell$ during one time unit after its creation:
\[
\eta_\mathrm{seen} = \frac{w_\mathrm{seen}(\ell)}{T - t(\ell)}, \quad \eta_\mathrm{repost} = \frac{w_\mathrm{repost}(\ell)}{T - t(\ell)}
\]
where $w(\ell)$ is the number of messages seen or reposted through $\ell$; $t(\ell)$ is the time when $\ell$ was created; and $T$ is the time of the last action recorded in our dataset.
Both seen and reposted messages are considered, as they represent different types of traffic; the former are what is visible to a user, and the latter are what a user is willing to share. We compute the link efficiency of every grandparent, origin, and triadic closure link.
As shown in Fig.~\ref{fig:eff}, both grandparent and origin links exhibit higher efficiency than triadic closure links, irrespective of the type of traffic. By shortening the paths of information flows, more posts from the content generators reach the consumers.

\section{Rules of Network Evolution}
\label{sec:5}

To infer the different link creation strategies from the observed data, we characterize users with a set of probabilities associated with different actions, and approximate these parameters by \linebreak Maximum-Likelihood Estimation (MLE)~\cite{cowan}. For each link $\ell$, we know the actual creator and the target; we can thus compute the likelihood $f(\ell | \Gamma, \Theta)$ of the target being followed by the creator according to a particular strategy $\Gamma$, given the network configuration $\Theta$ at the time when $\ell$ is created.
The likelihoods associated with different strategies can be mixed according to the parameters to obtain a model of link creation behavior.
Finally, assuming that link creation events are independent, we can derive the likelihood of obtaining the empirical network from the model by the product of likelihoods associated with every link. The higher the value of the likelihood function, the more \emph{accurate} the model.


\subsection{Single Strategies}

Let us consider five link creation mechanisms and their combinations: 
\begin{description}
\item[Random (Rand):] follow a randomly selected user who is not yet followed. 
\item[Triadic closure ($\Delta$):] follow a randomly selected triadic node.
\item[Grandparent ($G$):] follow a randomly selected grandparent.
\item[Origin ($O$):] follow a randomly selected origin.
\item[Traffic shortcut ($G \cup O$):] follow a randomly selected grandparent or origin.
\end{description}
Other mechanisms for link creation could be similarly incorporated, such as social balance~\cite{easley2010networks} and preferential attachment~\cite{BA}. However, preferential attachment is built on the assumption that everyone knows the global connectivity of everyone else, which is not realistic. The strategies considered here essentially reproduce and extend the copy model~\cite{kumar2000stochastic}, approximating preferential attachment with only local knowledge.

To model link creation with a single strategy, we can use a parameter $p$ for the probability of using that strategy, while a random user is followed with probability $1-p$. 
The calculation of maximum likelihood, taking the single strategy of grandparents as an example, is as follows:
\begin{align*}
\mathcal{L}_{G}(p)
	&=  \prod_{\ell=1}^{L} \left( pf(\ell | G, \Theta) + (1-p)f(\ell | \text{Rand}, \Theta) \right)\\
	&= \prod_{\ell=1}^{L} \left( p\frac{\mathbf{1}_{G}(\ell)}{N_G(\ell)} + (1-p)\frac{1}{N(\ell) - k(\ell) - 1} \right)\\
	&= \prod_{\mathbf{1}_{G}(\ell)=1} \left( \frac{p}{N_G(\ell)} + \frac{1-p}{N(\ell)- k(\ell) - 1} \right) \\
	& \quad \prod_{\mathbf{1}_{G}(\ell)=0} \frac{1-p}{N(\ell) - k(\ell) - 1}. 
\end{align*}
Note that since a follow action can be ascribed to multiple strategies, it can contribute to multiple terms in the log-likelihood expression. For instance, a link could be counted in both $f(\ell | G, \Theta)$ and $f(\ell | \text{Rand}, \Theta)$.
For numerically stable computation, we maximize the log-likelihood:
\begin{multline*}
\log \mathcal{L}_{G}(p)
	= \sum_{\mathbf{1}_{G}(\ell)=1} \ln \left( \frac{p}{N_G(\ell)} + \frac{1-p}{N(\ell) - k(\ell) - 1} \right) + \\
	\sum_{\mathbf{1}_{G}(\ell)=0} \ln \frac{1-p}{N(\ell) - k(\ell) - 1}.
\end{multline*}
Similar expressions of log-likelihood can be obtained for other strategies ($\Delta$, $O$, and $G \cup O$).

\begin{figure*}
\centering
\includegraphics[width=\textwidth]{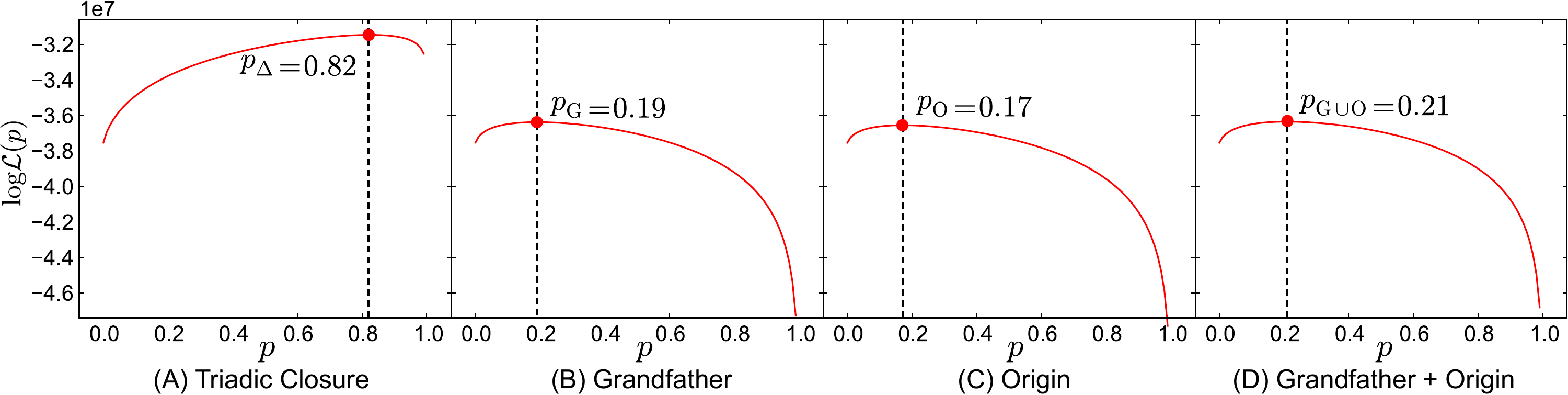}
\vspace{-1em}
\caption{
Plot of the log-likelihood $\log \mathcal{L}(p)$  as a function of link creation strategy probabilities for models with a single strategy. The red circles mark the maximized $\log \mathcal{L}(p)$. }
\vspace{-1em}
\label{fig:model1}
\end{figure*}

It is not trivial to obtain the best $p$ analytically, so we explore the values of $p \in (0,1)$ numerically (Fig.~\ref{fig:model1}). Triadic closure dominates as a single strategy, with $p_\mathrm{\Delta}=0.82$, consistently with the large number of triadic closure links observed in the data. Traffic-based strategies alone account for about 20\% of the links.

\subsection{Combined Strategies}

For a more realistic model of the empirical data, let us consider combined strategies with both triadic closure and traffic-based shortcuts. For each link $\ell$, the follower with probability $p_1$ creates a shortcut by linking to a grandparent ($G$), an origin ($O$), or either of them ($G \cup O$); with probability $p_2$ follows a triadic node ($\Delta$); and with probability $1-p_1-p_2$ connects to a random node.

Taking the combined strategy with grandparent as an example, we compute the log-likelihood as:
\begin{flalign*}
&\log \mathcal{L}_{G+\Delta}(p_1,p_2) \\
& = \log \prod_{\ell=1}^{L} [ p_1 f(\ell | G, \Theta) + p_2 f(\ell | \Delta, \Theta) \\
& \quad +  (1-p_1-p_2) f(\ell | \text{Rand}, \Theta) ] \\
& =  \sum_{\substack{\mathbf{1}_G(\ell)=1\\ \mathbf{1}_{\Delta}(\ell)=1}} \log \left(\frac{p_1}{N_G(\ell)} + \frac{p_2}{N_{\Delta}(\ell)} + \frac{1-p_1-p_2}{N(\ell) - k(\ell) - 1} \right) \\
& \quad  + \sum_{\substack{\mathbf{1}_G(\ell)=1\\ \mathbf{1}_{\Delta}(\ell)=0}}  \log \left(\frac{p_1}{N_G(\ell)} + \frac{1-p_1-p_2}{N(\ell) - k(\ell) - 1} \right) \\
& \quad  +  \sum_{\substack{\mathbf{1}_G(\ell)=0\\ \mathbf{1}_{\Delta}(\ell)=1}} \log \left(\frac{p_2}{N_{\Delta}(\ell)} + \frac{1-p_1-p_2}{N(\ell) - k(\ell) - 1} \right) \\
& \quad  + \sum_{\substack{\mathbf{1}_G(\ell)=0\\ \mathbf{1}_{\Delta}(\ell)=0}} \log \frac{1-p_1-p_2}{N(\ell) - k(\ell) - 1}.
\end{flalign*}
Once again, many follow actions can create both triadic closure links and traffic shortcuts, so they can contribute to multiple terms in the log-likelihood expression.

\begin{figure}[!b]
\centering
\includegraphics[width=0.975\columnwidth]{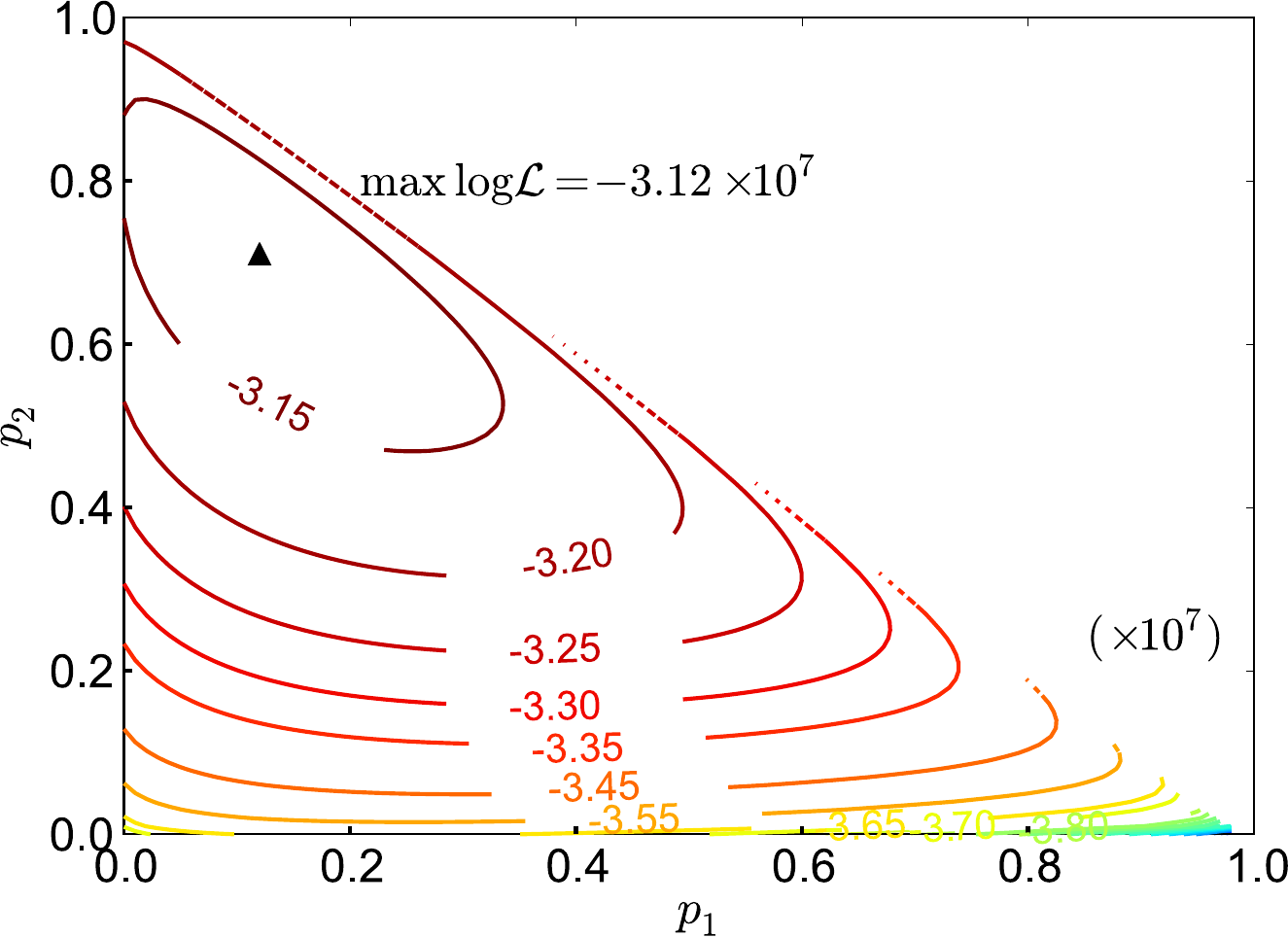}
\vspace{-1em}
\caption{The contour plot of log-likelihood $\log \mathcal{L}(p_1, p_2)$ for the combined strategy of creating traffic shortcuts ($G \cup O$) with probability $p_1$ and triadic closure links ($\Delta$) with probability $p_2$. The black triangle marks the optimum.}
\label{fig:model12}
\end{figure}

It is hard to obtain the optimal solution analytically. We numerically explore the values of $p_1$ and $p_2$ in the unit square to maximize the log-likelihood. The best combined strategy is the one considering both grandparents and origins as well as triadic closure (see Fig.~\ref{fig:model12}).
The parameter settings and the maximum likelihood values for all tested models are listed in Table~\ref{table:models}. We can compare the quality of these models by comparing their maximized $\log \mathcal{L}$'s. The combined models with both traffic shortcuts and triadic closure yield the best accuracy. In these models, triadic closure accounts for 71\% of the links, grandparents and origins for 12\%, and the rest are created at random.

Thus far we have assumed that each user has the same behavior; in the next section we model each user separately.

\begin{table}
\caption{The best parameters in different models and corresponding values of maximized log-likelihood function.}
\centering
\begin{tabular}{c  c  c  c}
\hline
Strategy & Model & Parameters & $\max\log\mathcal{L}$ \\
\hline\hline
Baseline & $\mathrm{Rand}$ & -- & $-3.75\times 10^7$ \\
\hline
\multirow{4}{*}{Single}
	& $\Delta$ & $p=0.82$ & $-3.15\times 10^7$ \\ 
	& $G$ & $p=0.19$ &  $-3.64\times 10^7$ \\ 
	& $\mathrm{O}$ & $p=0.17$ &  $-3.65\times 10^7$ \\ 
	& $G\cup\mathrm{O}$ & $p=0.21$ &  $-3.63\times 10^7$ \\
\hline
\multirow{6}{*}{Combined}
	& \multirow{2}{*}{$G+\Delta$} & $p_1=0.12$ & \multirow{2}{*}{$-3.12\times 10^7$} \\ 
	& & $p_2=0.71$ & \\
	& \multirow{2}{*}{$\mathrm{O}+\Delta$} & $p_1=0.10$ & \multirow{2}{*}{$-3.13\times 10^7$} \\
	& & $p_2=0.73$ & \\
	& \multirow{2}{*}{$G\cup\mathrm{O}+\Delta$} & $p_1=0.12$ & \multirow{2}{*}{$-3.12\times 10^7$} \\
	& & $p_2=0.71$ & \\
\hline
\end{tabular}
\vspace{-1em}
\label{table:models}
\end{table}

\section{User Behavior}
\label{sec:6}

The MLE models for describing the system behavior can be similarly employed to characterize the strategy of an individual user. 
Let us focus on the model $G \cup O +\Delta$ that best reproduces the empirical data at the global level. We run MLE to explain the links created by each user independently. We consider users with at least 20 in-links, such that MLE is meaningful.
For easier interpretation, let us call $p_\mathrm{traffic} = p_1$, $p_\mathrm{structure} = p_2$ and $p_\mathrm{random} = 1-p_1-p_2$. Each user has her own set of parameters.

\subsection{User Strategy Classification}

Using the Expectation-Maximization (EM) algorithm~\cite{celeux1992classification,dempster1977maximum}, users are clustered into several classes based on $p_\mathrm{traffic}$, $p_\mathrm{structure}$ and $p_\mathrm{random}$. 
EM iteratively preforms an expectation step to compute the probability that each instance belongs to each class, and a maximization step in which latent variables of classes are altered to maximize the expected likelihood of the observed data. EM decides how many clusters to create by cross validation. This procedure yields five classes:

\begin{description}

\item[\textbf{Information-Oriented} (Info):] 
People prefer to follow someone \linebreak from whom or through whom they have received messages.

\item[\textbf{Friend of a Friend} (Friend):]
People follow users two steps away to form triadic closure, almost exclusively.

\item[\textbf{Casual Friendship} (CFrd):]
People tend to follow a set of users their friends are following; they also link to random users occasionally.

\item[\textbf{Mixture} (Mix):]
Miscellaneous behavior of creating traffic shortcuts, connecting others by triadic closure, and following random people. 

\item[\textbf{Random Browsing} (Rand):]
People have a much higher preference for following a random user who is not close in either the follower or the message flow network. 
``Random'' does not necessarily imply the absence of any rule; there can be other strategies not explored in our model, i.e., following a celebrity on purpose (similar to preferential attachment).

\end{description}

\begin{figure}
\centering
\includegraphics[width=\columnwidth]{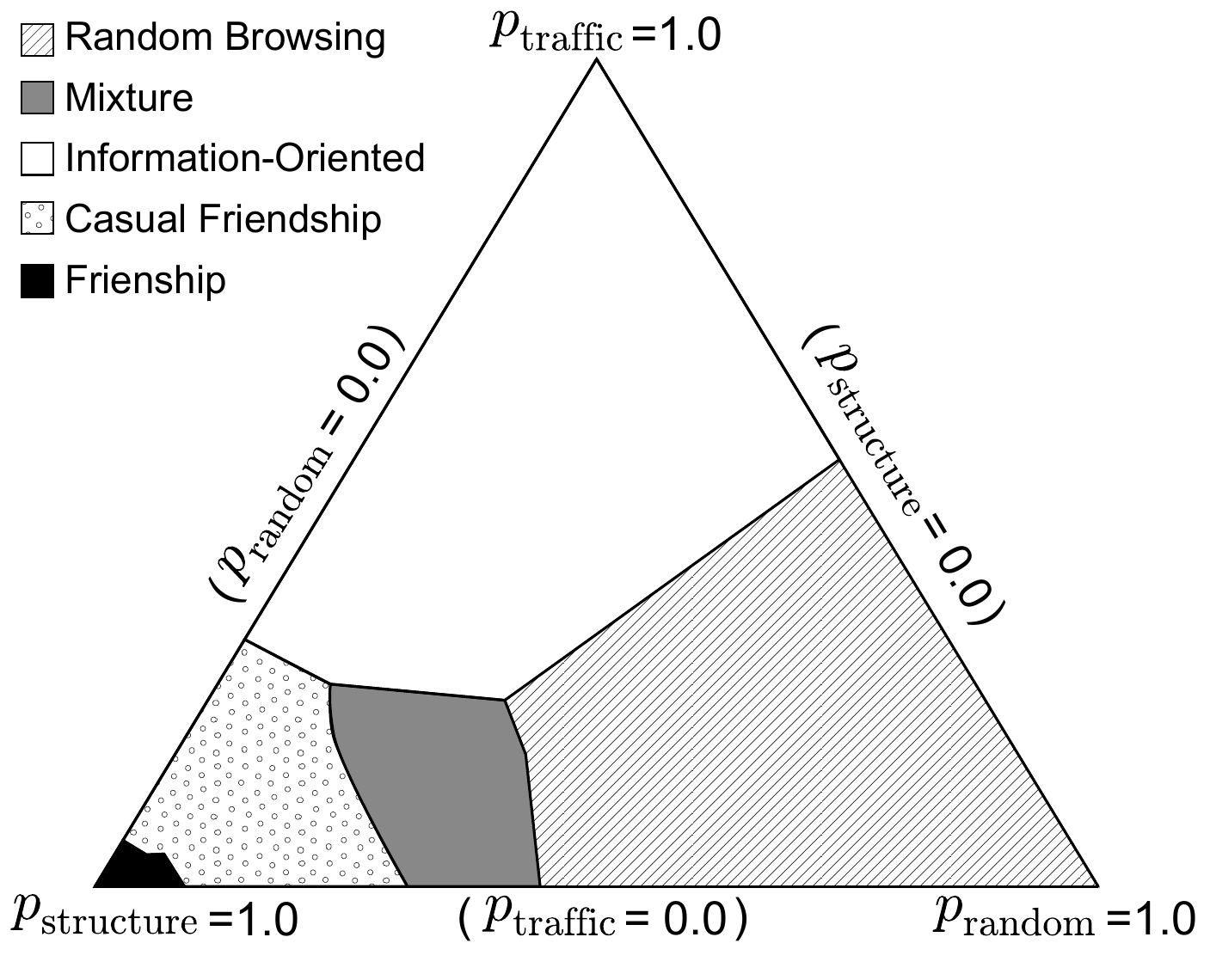}
\vspace{-1em}
\caption{Ternary plot of users according to $p_\mathrm{traffic}$, $p_\mathrm{structure}$ and $p_\mathrm{traffic}$.}
\label{fig:behavior_ternary}
\end{figure}

Table~\ref{table:behavior_class} displays the parameter averages for users in each class, representing the overall behavior pattern in that class.
Users in the mixture category behave similarly to the average across all users.
Fig.~\ref{fig:behavior_ternary} illustrates how users in different classes are mapped into the parameter space with the probability of each link creation strategy as one dimension.

\begin{table}
\caption{Classes of user link creation strategy}
\begin{center}
\begin{tabular}{c  c  c   c  c}
\hline
Class & \#Users  & $\langle p_\mathrm{traffic}\rangle$ & $\langle p_\mathrm{structure}\rangle$ &  $\langle p_\mathrm{random}\rangle$ \\
\hline
All Users & 45,708 & 0.07 & 0.77  & 0.17 \\
Info & 4,750 & 0.52  & 0.36 & 0.13 \\
Friend & 12,797 & 0.00 & 0.96 & 0.04 \\
CFrd & 23,469 & 0.01 & 0.80 & 0.19 \\
Mix & 2,524 & 0.07 & 0.63 & 0.30 \\
Rand & 2,168 & 0.09 & 0.32 & 0.59 \\
\hline
\end{tabular}
\vspace{-2em}
\end{center}
\label{table:behavior_class}
\end{table}%

\subsection{Characterization of User Classes}

\begin{figure*}[t]
\centering
\includegraphics[width=\textwidth]{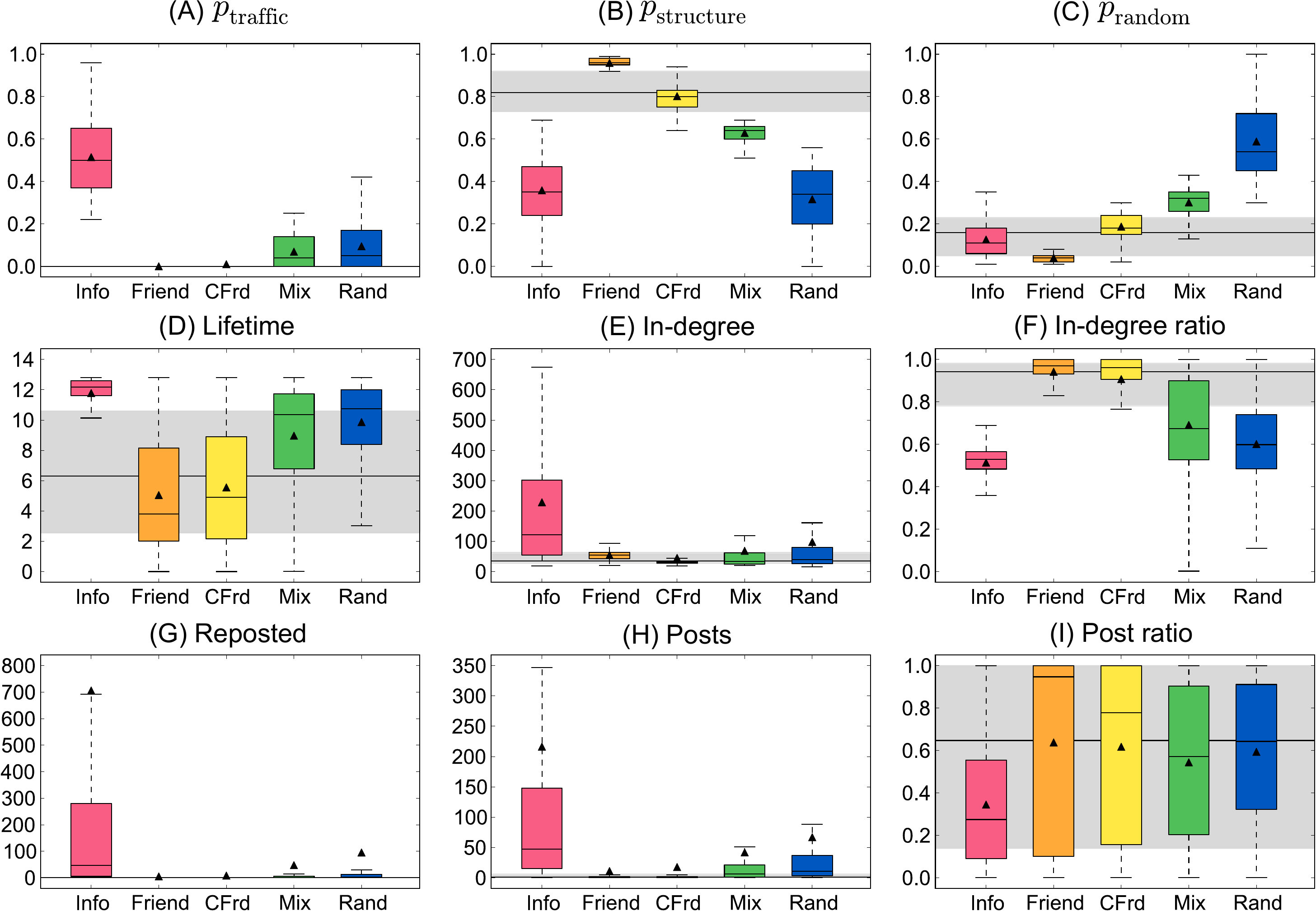}
\vspace{-1em}
\caption{Various features of users in different classes. 
The lifetime of a user is measured by how many others join the system after him.
The in-degree is the number of people a user is following, $k$, and the in-degree ratio is $k/(k+k_{out})$ where $k_{out}$ is the number of followers.
``Reposted'' refers to the number of times that a user's messages are reposted by others. ``Posts'' denotes the number of messages generated by a user excluding reposts. The post ratio is the fraction of all posts by a user (including reposts) that are originated by that user.
Each box shows data within lower and upper quartile. Whiskers represent the 99th percentile. The triangle and line in a box represent the median and mean, respectively. The grey area and the black line across the entire figure mark the interquartile range and the median of the measure across all links, respectively.
}
\label{fig:behavior_boxplot}
\end{figure*}

To further differentiate users with different link creation strategies, let us look at several structural and behavioral characteristics of each class. Figs.~\ref{fig:behavior_boxplot}(A-C) show how users in different classes create social links by comparing $p_\mathrm{traffic}$, $p_\mathrm{structure}$ and $p_\mathrm{random}$.

As shown in Fig.~\ref{fig:behavior_boxplot}(D), information-oriented users have been active longer than users in friendship classes.
Similarly, information-oriented users tend to follow more people (Fig.~\ref{fig:behavior_boxplot}(E)). Information-oriented users have even more followers compared to friendship-driven users (Fig.~\ref{fig:behavior_boxplot}(F)). This suggests that they tend to be more influential, as confirmed by considering the number of times that their messages are reposted (Fig.~\ref{fig:behavior_boxplot}(G)). Friendship-driven users follow a few people while essentially nobody is following them. Such a passive role can be explained by their short lifetime. All of these results are consistent with Fig.~\ref{fig:preference}. 

Finally, Figs.~\ref{fig:behavior_boxplot}(H-I) suggest that, while information-oriented users tend to produce more messages, their role is more that of spreaders than producers of information compared to other classes. 

In this analysis, the parameters are fit according to the entire lifetime of each user. Focusing on the first 20 or 50 links does not yield qualitatively different results.

\section{Conclusion}

The study of the feedback loop between the dynamics \emph{of} and \emph{on} the network --- how the network grows and how the information spreads --- offers a promising framework for understanding social influence, user behavior, and network efficiency in the context of micro-blogging systems.

The results presented in this paper show that while triadic closure is the dominant mechanism for social network evolution, it is mainly relevant in the early stages of a user's lifetime. As time progresses, the traffic generated by the dynamics of information flow on the network becomes an indispensable component for user linking behavior. As users become more active and influential, their links create shortcuts that make the spread of information more efficient in the network. Users whose following behavior is driven by the information they see are a minority of the population, but play a key role in the information diffusion process. They produce more information, but, even more importantly, they act as spreaders of the information they collect widely across the network.

While existing link prediction algorithms~\cite{LibenNowell2007LinkPred,Clauset:2008fk,backstrom2011supervised,lou2010,Schifanella20120folk} are not designed to explain the network evolution in a dynamic setting, the MLE framework could in principle be used to assess which link prediction methods are more consistent with the longitudinal structural changes observed in the network, by treating the prediction at each step as a link creation strategy. These approaches will be explored in future work.

We believe our findings apply generally to techno-social networks, and in particular information diffusion networks and (micro)blogs. Analyses of other micro-blogging systems, such as Twitter and Weibo, would be needed to confirm this, but will be challenging due to the difficulty of obtaining full longitudinal data about user actions on the social network.


\vspace{1em}
\noindent \textbf{Acknowledgments.} This project is a collaboration between the Center for Complex Networks and Systems Research (\url{cents.indiana.edu}) at Indiana Univ. and Yahoo! Research Barcelona. During the project, CC was at Yahoo!; NP and BG were at CNetS; RS was visiting CNetS; and JR was visiting Yahoo! as a PhD student at CNetS.
We are grateful to Alessandro Vespignani for helpful discussion, to Ricardo Baeza-Yates for support and access to the data, and to anonymous referees for valuable comments. 
This project was funded in part by the James S. McDonnell Foundation, NSF Grant No. 1101743, DARPA Grant No. W911NF-12-1-0037, and the Indiana Univ. School of Informatics and Computing.

\bibliographystyle{abbrv}
\bibliography{references}

\end{document}